# Comment: Expert Elicitation for Reliable System Design

**Andrew Koehler**

This article successfully identifies and addresses some of the most important challenges in the use of elicitation as part of engineered systems analysis. The authors make two key advances to the field. The first is that they perform an exhaustive synthesis of the probability elicitation literature relevant to the engineered systems context. The second advance to the field offered by this paper is that it reframes the elicitation literature around the limits, possibilities and actual constraints posed by systems engineering practice. This latter point is no small contribution—the authors have successfully opened a much needed discussion as to why system elicitation differs fundamentally from cultural ethnography methods, and why risk estimation and "systems ethnography" (elicitation of system dependencies and evolving uncertainties) are only partially informed by methods developed for identified and stable single distribution elicitation.

I like the overview of the systems engineering (SE) life cycle and the link made to reliability through the $r = r(d, p, u, m, c)$ relationship. In a follow-up paper or discussion it would be interesting to learn more about the types of systems the authors have studied. In thinking about how to elicit system information over a complete range of engineering efforts, it quickly becomes apparent how hard it is to characterize the elicitation effort (and why this article is such a notable exception to the general lack of disciplined study of this qualitative field). Because the elicitation problem varies greatly depending on the specific form of a technical system, as well as local analytic and decision-making realities, perhaps what this article has accomplished is to identify a core set of considerations for reliability elicitation. One could imagine additions to the core set of issues developed in this paper that could lead to a kind of technical system elicitation taxonomy.

Much of the article discusses systems characterized by a "closed loop" or "spiral" type of systems engineering process. In many cases, such as systems manufactured in multiple batches, this closed-loop model is accurate. In some cases, however, the SE process is not closed loop in form and while the basic building blocks of the elicitation task identified in this article remain valid, additional challenges can emerge. The bulk of my exposure to the use of elicitation methods as part of reliability prediction has involved either weapon or long lead facility construction systems. These systems have traditionally either been developed using a "waterfall" SE model or been deployed for a sufficiently long time that design and fabrication of new versions have ended. In the case of waterfall system engineering projects, there can be a great divide between design and operational life cycle phases, and often relatively little system knowledge (especially tacit knowledge) is transferred between the communities involved with each phase. By the time these kinds of systems are deployed, it is not uncommon that the design team has largely been scattered, downsized or otherwise dispersed. Because validation information generated during the operation of such systems cannot be passed back to the entity responsible for design, perhaps a valid extension of the authors' elicitation closed loop to the waterfall case taxonomy might include two additional types of expert knowledge:

1. System reliability predictions, associated uncertainties and estimates for component behavior may be dependent on changes in expertise and team composition between life cycle phases, in system models used by operators, in operational constraint shifts or in system program importance. For example, the military may transfer an operational, but no longer produced, weapon from front-line troops to support units, and this may entirely change the nature of reliability concerns, the amount of testing performed or the nature of operational evaluation. This lack of continuity in expertise over a waterfall system's life cycle implies it may be necessary to understand how system knowledge is changing between system engineering life cycle time periods.

*Andrew Koehler is a Staff Member, Statistical Sciences Group, MS F600, Los Alamos National Laboratory, Los Alamos, New Mexico 87545, USA e-mail: akoehler@lanl.gov.*

2. A divide between design and operational stages may also make it necessary to understand whether the system's operational history has been disrupted by (what I call) forward casting. By "forward casting" I refer to situations where operators employ a waterfall system in ways that differ from the operational assumptions made at the system's birth. The decision to operate an existing system, designed as part of a waterfall SE process, outside of its requirement environment is essentially the same as reaching back into the system's life cycle, virtually changing a design requirement decision (often unilaterally and without any input from a perhaps no longer existing design authority) and propagating the changed set of requirements onto a system possibly incapable of adapting. Because in such waterfall system cases the design activity has ended (as opposed to a more spiral SE effort), a considerable degree of system management uncertainty, and hence opportunity for expert elicitation, is generated in the case when "forward casts" are being contemplated. Predictions for system outcome as a result of forward casting are typically resolved by reliance on expert and subjective judgment, because more quantitative models are unavailable. In some cases, sufficient design records may not even exist to characterize how requirements drove existing configuration choice, never mind to predict what may happen if the system's operational patterns are fundamentally changed.

Thinking about the elicitation problem characterized by these two factors (lack of expertise continuity and the problem of forward casting requirements for an existing system), we see that many of the issues identified in this article remain true. It strengthens the excellent points made in the article that many of the elicitation issues raised become more complicated as system elicitation characteristics differ from the spiral SE model. However the importance of understanding the links between how a system is built, how the system works and how qualitative information about it can be gathered and used for quantitative reliability estimation remains essential.

Beyond characterizing relationships between the systems engineering task and reliability elicitation, there are many other promising avenues for research made possible by the paper's successful linkage of presently disparate bodies of knowledge. For example, the paper performs a service by distilling, out of a large disjoint literature, four main elicitation roles that play an important part in determining a complex engineered system information gathering strategy. This provides a solid foundation for additional thinking about how such roles relate to the problem of keeping track of a subject's span of expertise. The conditional nature of the expert plays a major role in system elicitation, far more so than is common in the case of more traditional ethnographic settings. This poses different problems than just that of conflicting opinions, because before an information combination strategy can be employed to resolve expert disagreement, a determination must be made that these expert estimates are informative about a comparable metric. At present, the elicitation and engineering fields have not engaged this issue adequately.

Additionally, limits on an expert's span of control in the case of complex engineered systems imply that no single system participant can verify the entire structure of the system. Given the immensely large dependency structures needed to describe the performance of a system or even a system of systems, it can by no means be certain that important parts or relationships are being captured. The seemingly continual emergence of Murphy's law and other unexpected system behaviors proves the difficulty of characterizing what factors determine a complex system's performance. While the literature has focussed on how to gather and combine estimates about an identified distribution, much less attention has been paid to understanding the overall structure of variables that define the system and their interrelationships. This is an epistemic problem that is not resolvable at the level of combining different estimates about a single distribution. As systems of increasing behavioral complexity are deployed, we may be moving toward a state where creation of a "coherent" system model for prediction may depend on maintaining multiple ultimately nonreducible versions of subsystem structures. Instead of arriving at *a* model, we may be forced by some systems of systems to admit that a level of complexity has been reached where no absolute, even momentarily static, coherent structure can be resolved. In such cases, reliability estimates may come to be based on multiple, simultaneous models for the system structure—with a joint reliability distribution resulting from some combination of these models. This connects partially to the article's discussion on expert pooling; however, it will present a number of verification challenges that differ from the problem of parameter estimate combination.

As structured system elicitation and analysis methods become a more important part of systems analysis and prediction, it is a great challenge to the literature to build on what is known and to understand when additional challenges can be expected. This article takes the field a great step forward by making the link between the characteristics of a technology, the SE process, and the fundamental issues in choice of an elicitation strategy.